\documentclass[twocolumn]{cinc}
\usepackage{graphicx}
\usepackage{comment}
\newcommand{\heading}[1]{\textbf{#1}}

\usepackage{amsmath}
\usepackage{textcomp} 

\usepackage{xcolor}

\begin{document}
\bibliographystyle{cinc}

\title{CardioLab: Laboratory Values Estimation from Electrocardiogram Features - An Exploratory Study}



\author {Juan Miguel Lopez Alcaraz$^{1}$, Nils Strodthoff$^{1}$ \\
\ \\ 
 $^1$ AI4Health Division, Carl von Ossietzky Universität Oldenburg, Oldenburg, Germany.}

\maketitle

\begin{abstract}
Laboratory value represents a cornerstone of medical diagnostics, but suffers from slow turnaround times, and high costs and only provides information about a single point in time. The continuous estimation of laboratory values from non-invasive data such as electrocardiogram (ECG) would therefore mark a significant frontier in healthcare monitoring. Despite its potential, this domain remains relatively underexplored. In this preliminary study, we used a publicly available dataset (MIMIC-IV-ECG) to investigate the feasibility of inferring laboratory values from ECG features and patient demographics using tree-based models (XGBoost). We define the prediction task as a binary problem of whether the lab value falls into low or high abnormalities. We assessed model performance with AUROC. Our findings demonstrate promising results in the estimation of laboratory values related to different organ systems. While further research and validation are warranted to fully assess the clinical utility and generalizability of the approach, our findings lay the groundwork for future investigations for laboratory value estimation using ECG data. Such advancements hold promise for revolutionizing predictive healthcare applications, offering faster, non-invasive, and more affordable means of patient monitoring.
\end{abstract}

\section{Introduction}

\heading{Clinical relevance} Abnormal laboratory values play a crucial role in clinical settings as they indicate underlying health conditions, sometimes posing severe and potentially life-threatening risks, especially in cases of serious illnesses \cite{fan2016diagnosing,lim2015risk}. These abnormal values significantly increase the risk of mortality compared to normal ranges \cite{dzankic2001prevalence}, highlighting the need for vigilant monitoring and effective management strategies. Continuous monitoring of patients often involves inherent limitations such as invasive procedures like frequent blood sampling, which are resource-intensive \cite{beliveau2018decreasing} and suffer from delays between sampling and obtaining results \cite{ezzie2007laboratory} or potentially missing staffing of laboratory medicine during the night, thereby limiting real-time monitoring capabilities.

\begin{table*}[htbp]
\caption{\label{tab:font} Performance results for the top 25 laboratory values presented by value name, abnormality setting, threshold defining abnormality by the number of units, (units), number of samples, class prevalence, and AUROC with 95\% confidence intervals.}
\vspace{4 mm}
\centering
\begin{tabular}{lcccc} \hline\hline
\textbf{Value}  & \textbf{Threshold} & \textbf{Unit} & \textbf{Samples [Prev.]} & \textbf{AUROC (95\% CI)} \\ \hline
Albumin & 	\textgreater{}5.2 & g/dL & 57040[0.4\%] & 0.901 (0.901, 0.906) \\
Hemoglobin & 	\textgreater{}17.5 & g/dL & 236802[0.34\%] & 0.881 (0.881, 0.884) \\
NTproBNP & 	\textgreater{}353.0 & pg/mL & 23206[77.72\%] & 0.866 (0.865, 0.867) \\
Acetaminophen & 	\textgreater{}30.0 & ug/mL & 1205[18.76\%] & 0.857 (0.856, 0.864) \\
Hematocrit & 	\textgreater{}51.0 & \% & 238567[0.58\%] & 0.847 (0.846, 0.849) \\
PT & \textless{}9.4 & sec & 137917[0.28\%] & 0.831 (0.828, 0.831) \\
Red Blood Cells & 	\textgreater{}6.1 & m/uL & 234079[0.31\%] & 0.818 (0.818, 0.822) \\
25-OH Vitamin D & \textless{}30.0 & ng/mL & 1101[55.4\%] & 0.815 (0.813, 0.822) \\
RDW-SD & \textless{}35.1 & fL & 73001[0.22\%] & 0.811 (0.811, 0.817) \\
INR(PT) & \textless{}0.9 & nan & 137973[0.22\%] & 0.803 (0.796, 0.801) \\
Urea Nitrogen & \textless{}6.0 & mg/dL & 240892[0.89\%] & 0.796 (0.796, 0.798) \\
Monocytes & \textgreater{}11.0 & \% & 171442[5.77\%] & 0.768 (0.768, 0.769) \\
Acetaminophen & \textless{}10.0 & ug/mL & 1205[48.22\%] & 0.768 (0.768, 0.775) \\
Absolute Basophil Count & \textless{}0.01 & K/uL & 51315[7.81\%] & 0.765 (0.765, 0.767) \\
Urea Nitrogen & \textgreater{}20.0 & mg/dL & 240892[41.6\%] & 0.756 (0.756, 0.756) \\
C-Reactive Protein & \textgreater{}5.0 & mg/L & 3520[59.86\%] & 0.753 (0.752, 0.757) \\
Cholesterol, HDL & 	\textless{}41.0 & mg/dL & 9023[23.31\%] & 0.749 (0.746, 0.749) \\
Bilirubin, Direct & \textgreater{}0.3 & mg/dL & 3438[57.24\%] & 0.748 (0.747, 0.752) \\
RDW-SD & \textgreater{}46.3 & fL & 73001[49.16\%] & 0.744 (0.744, 0.745) \\
Hemoglobin & \textless{}13.7 & g/dL & 236802[72.39\%] & 0.741 (0.741, 0.742) \\
Creatinine & 	\textgreater{}1.2 & mg/dL & 241968[28.52\%] & 0.738 (0.738, 0.739) \\
Sedimentation Rate & 	\textgreater{}20.0 & mm/hr & 1861[57.87\%] & 0.736 (0.734, 0.74) \\
pO2 & 	\textless{}85.0 & mm Hg & 35047[46.75\%] & 0.733 (0.732, 0.733) \\
Osmolality, Measured & 	\textless{}275.0 & mOsm/kg & 2784[26.22\%] & 0.729 (0.726, 0.732) \\
Bicarbonate & 	\textgreater{}32.0 & mEq/L & 231772[2.86\%] & 0.728 (0.728, 0.729) \\ \hline\hline
\end{tabular}
\label{table}
\end{table*}

\heading{Significance of abnormal laboratory values} In the United States, annual costs for preoperative laboratory testing alone amount to \$18 billion \cite{beliveau2018decreasing}. In intensive care units (ICUs), these costs can rise to \$14 billion, constituting up to 25\% of total ICU expenditures \cite{halpern2004critical}. Identifying abnormal lab values is crucial for patient care. In elderly surgical patients, preoperative tests often show abnormal creatinine (12\%), hemoglobin (10\%), and glucose (7\%) levels \cite{dzankic2001prevalence}. In ICUs, these tests are essential for adjusting drug dosages and procedures for critically ill patients, playing a key role in treatment strategies \cite{ezzie2007laboratory}.

\heading{Monitoring and management strategies} Current guidelines for monitoring patients with abnormal lab values recommend instruments and methods for early detection, continuous monitoring, and effective interventions. They include organ-specific tests, like liver function \cite{newsome2018guidelines}, and are adapted for settings such as the emergency department (ED) \cite{10.1373/49.3.357}. While traditional phlebotomy is common, non-invasive alternatives are emerging for biomarkers like glucose \cite{vashist2012non} and hemoglobin \cite{shah2014accuracy}.

\heading{Integration of ECG data for values prediction} The use of ECG data for laboratory values estimation has been briefly investigated, where previous works demonstrated predictive capabilities in glucose \cite{chiu2022utilization}, serum potassium \cite{Chiu2024.05.08.24307064}, and electrolyte imbalances \cite{kwon2021artificial,von2024evaluating}. Lastly, other predictive modalities such as vital signs from wearable sensors were investigated to predict laboratory values \cite{dunn2021wearable}.

\heading{Contribution} While previous research has explored the correlation between abnormal laboratory values and changes in electrocardiogram (ECG) readings \cite{surawicz1967relationship,nunez2009relationship}, the full potential of utilizing ECG data for accurately estimating these abnormalities remains largely untapped. Therefore, in this work, we propose a first exploratory study in which we investigate the feasibility of estimating laboratory values from ECG features with the addition of other data modalities such as patient demographics.

\section{Methods}

\heading{Dataset} The dataset comprises data sourced from the publicly available MIMIC-IV \cite{johnson2023mimic,goldberger2000physiobank} and MIMIC-IV-ECG \cite{MIMICIVECG2023,goldberger2000physiobank} datasets. It includes a comprehensive set of non-invasive features: demographics such as gender, age, and race, as well as vital signs such as temperature, heart rate, respiration rate, oxygen saturation, diastolic blood pressure, systolic blood pressure; and finally, ECG features including RR interval, P onset, P end, QRS onset, QRS end, and T end (all in milliseconds), as well as P axis, QRS axis, and T axis (in degrees). In terms of target abnormalities, we work on binary classification cases where we define a positive case when the considered laboratory value is lower or higher than the patient-wise median low or high threshold values (provided within MIMIC-IV) across all samples of the same laboratory value. For data sampling, the estimation task involves using the nearest vital signs recorded within a 30-minute interval from the ECG data to predict laboratory abnormalities. These predictions are based on the closest lab values within a 60-minute window.  For dataset splitting, we follow the patient-based stratification based on demographics and diagnoses proposed by MIMIC-IV-ECG-ICD \cite{strodthoff2024prospects}, which splits with a train, validation, and test ratio of 18:1:1. Finally, for this work, we consider final laboratory value cases where we obtain at least 10 positive and 10 negative cases per fold.

\heading{Models and performance evaluation} We fit and train individual extreme gradient boosting (XGBoost) tree models with a max depth of 1 per laboratory value in a binary classification setting. We evaluate performances based on areas under the respective receiver operating curves (AUROC). To assess statistical uncertainty resulting from the finite size and specific composition of the test set, we use empirical bootstrapping on the test set with $n = 1000$ iterations and report 95\% confidence intervals.

\section{Results}

\heading{Overall Predictive Performance} Table~\ref{table} contains the performance results in terms of AUROC for the 25 best-performing individual laboratory value abnormalities. This includes the threshold for their label definition, units, number of samples, and prevalence. This selection of laboratory values, all of which represent blood fluids, underscores the model's capacity to predict abnormalities in diverse bodily systems. Notable among these are values related to cardiovascular function (Albumin, Hemoglobin, NTproBNP), coagulation (PT, INR), and oxygen transport (pO2). The model also effectively addresses renal function (Urea Nitrogen, Creatinine), immune response (Monocytes, Absolute Basophil Count), and inflammatory processes (C-Reactive Protein, Sedimentation Rate). Moreover, it accurately identifies metabolic (Cholesterol, HDL, Bicarbonate, Osmolality), endocrine (25-OH Vitamin D), and hepatic (Bilirubin, Direct) abnormalities. The inclusion of values like Acetaminophen and RDW-SD highlights the model's versatility in managing both common and more specialized laboratory measurements.

\section{Discussion} 
\heading{Clinical significance} This work enhances ECG predictive capabilities beyond traditional applications, in line with its counterpart \cite{alcaraz2024estimationcardiacnoncardiacdiagnosis} that demonstrates the predictability in particular of non-cardiac conditions from ECG features and clinical metadata. The model's ability to predict a wide range of clinically significant laboratory abnormalities using only ECG features, demographics, and basic, non-invasive vital signs is a major advancement in healthcare. By identifying critical cardiovascular issues such as those related to albumin and hemoglobin levels, the model enables early detection and intervention without the need for expensive or invasive testing. Its capacity to predict heart failure risk through NTproBNP levels and identify coagulation abnormalities, such as those indicated by PT and INR, further demonstrates its potential to improve patient outcomes by facilitating timely and targeted care. Moreover, the model's ability to assess renal function, immune response, and inflammatory processes using easily obtainable data could significantly enhance diagnostic accuracy and accessibility, particularly in resource-limited settings. This approach not only streamlines clinical workflows but also broadens the scope of preventive care, making it easier to monitor and manage patient's health effectively while providing faster and low-cost results.

\heading{Limitations} We identify several limitations in our current work. Firstly, for clinical consideration, the present promising results will have to be externally validated based on a separate population cohort. Secondly, we estimate the lab value with the closest time difference to the point in time where the ECG was taken. Using a fixed intervals between features and targets for all samples might mitigate variance. Thirdly, our work builds on constant lower and higher abnormality thresholds extracted from MIMIC. More reliable threshold choices, also depending on patient characteristics such as age and gender, should be investigated in the future.

\heading{Future work}

Firstly, our proposed method enables an easy integration of explainable methods such as Shapley additive explanations (SHAP) \cite{lundberg2020local} which will allow us to identify the specific set of features that contributing most to abnormality predictions. This would allow to quantify the relative impact of the different input modalities, demographics, vital signs, or ECG features on the prediction performance.

Secondly, in addition to the assessment of the overall model performance, it will be very instructive to also investigate the performance on specific subgroups defined for example based on demographics such as gender, ethnicity, and age groups. However, one must consider that certain groups might suffer from biased results such as abnormal values being considered normal at certain aging stages, which goes in line with more finegrained threshold definitions as discussed above.

Thirdly, while this work investigates the predictive performance of laboratory values by discriminating ECG features within a largely critically ill, it might be informative to investigate the same question in a healthy subgroup. This parallels recent efforts to characterize age-related ECG changes in a healthy cohort \cite{ott2024using}. 

Finally, previous research has demonstrated the superiority of using ECG waveforms over ECG features for diagnosis and decompensation prediction \cite{alcaraz2024mdsedmultimodaldecisionsupport}. Extending this work to include waveforms rather than relying solely on tabular data would be a valuable direction for future research.

\heading{Data and code availability} Code for dataset preprocessing and experimental replications can be found in our dedicated repository \cite{githubAI4HealthUOLCardioLab}.

\bibliography{refs}

\begin{correspondence}
Nils Strodthoff \\
Carl von Ossietzky Universität Oldenburg
Fakultät VI - Medizin und Gesundheitswissenschaften
Department für Versorgungsforschung
Abteilung AI4Health
Ammerländer Heerstr. 114-118
26129 Oldenburg, Deutschland \\
nils.strodthoff@uol.de
\end{correspondence}

\end{document}